# Fractionally Charged Particles in Cosmic Rays

George Bashindzhagyan, Natalia Korotkova

*Skobeltsyn Institute of Nuclear Physics Lomonosov Moscow State University, Moscow 119991, Russia*

**Abstract**: The results of many experiments on a search of fractionally charged particles in cosmic rays have been reviewed. The registered by ATIC and PAMELA experiments change of the proton energy spectrum at about 250 GeV can be explained if fractionally charged particles with another energy spectrum slope actually mixed with protons but cannot be separated because of a strong dE/dx fluctuations. The performed simulations show that multilayer detectors can seriously help in such separation. In the Aragats experiment performed using multilayer proportional counter combined with hadron calorimeter a group of 4e/3 like events with unexpectedly high average energy has been registered. It could be explained by their different from regular hadrons energy spectrum. The ATIC experiment ionization spectrum in single charged particle area has been examined. An interesting bump in 2e/3 charge region was observed. The events in the bump have very different from regular protons angular distribution.

## 1. Introduction

Many experiments have been performed to search for fractionally charged particles (FCP) in cosmic rays and various methods were used, for example [1-7].

The simple method to separate FCP from regular single charged ones is to use the fact that ionization generated by charged particles in gas or solid material like silicon quadratically depends on their charge. But serious problem appears because of so called Landau fluctuations of ionization, which has a typical half width about 35-45% for silicon detectors and 70-100% for gas detectors. It's also asymmetric with long tail, which makes even more difficult to identify particles with charge more than 1e. Therefore relatively good separation can be achieved if neighbor particles have relatively identical flux. But most of researchers expect that FCP flux has to be much lower than of single charged ones.

Serious improvement can be achieved if a few layers of detectors are used. Figure 1 shows the difference in separation for a single layer of typical silicon pad detector (left plot) and a system of six layers of such detectors (right plot). For simulation 200 GeV protons have been mixed with 4% of 2e/3 quarks, 4% of 4e/3 diquarks and 25% of alpha particles. One can see that not only 2e/3 but also 4e/3 will be well separated with six layers of silicon.

Promising version of universal multilayer device had been proposed for balloon or satellite application [8].

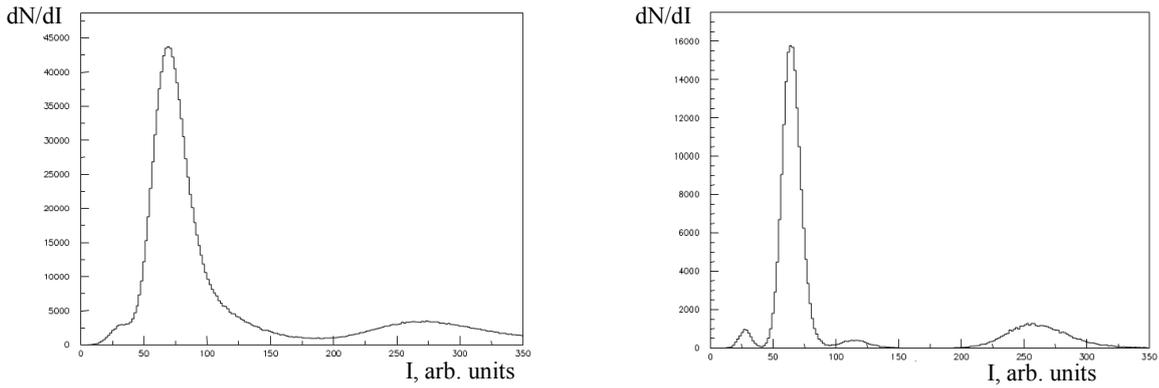

**Fig. 1.** Ionization spectrum for 200 GeV protons (100%) with admixture of alpha particles (25%) and particles with charge 2/3 and 4/3 e (4% and 4%). Signal to noise ratio is 10. Left plot: one layer of silicon detectors is used. Right plot: six layers are used.

## 2. Aragats experiment

This experiment [9] is unusual because 24 layers of gas proportional counter (PC) for precise charge measurements have been combined with hadron calorimeter and a hodoscope (Fig. 2). The experiment was performed at Aragats cosmic ray station (Armenia) at altitude of 3200 m above sea-level. The system was triggered by the calorimeter if hadron energy was above 80 GeV. Only single hadron type particles without accompanying air shower have been analyzed. 0.6 GeV muon ionization has been also measured. Muons were used for proportional counter calibration and for keeping PC gas amplification stable.

Two plots on Fig. 3 show muon and hadron ionization distributions. One can see that hadrons create ~1.5 times higher ionization because of Relativistic Rise of Ionization (RRI) in gases. A calibration curve obtained by fitting muon distribution has been used in the analysis of the hadron events and to separate a group of events, which had a considerably higher ionization.

The hadrons inside the calibration curve are a mixture of protons and pions. It means that they are on the plateau of the RRI curve and none of single charged hadrons may have higher ionization. But 14 events above this level have been observed. The probability that this is a random deviation is $5 \cdot 10^{-4}$. We cannot exactly evaluate their charge because it depends on their mass according to the RRI. If, for example, we suppose that they are 4e/3 diquarks their mass could be estimated as 30 GeV. But the most interesting and intriguing fact is that the average energy of these 14 events is 355 GeV while it's only 205 GeV for regular single charged hadrons. The probability that it is a random deviation is less than 0.01.

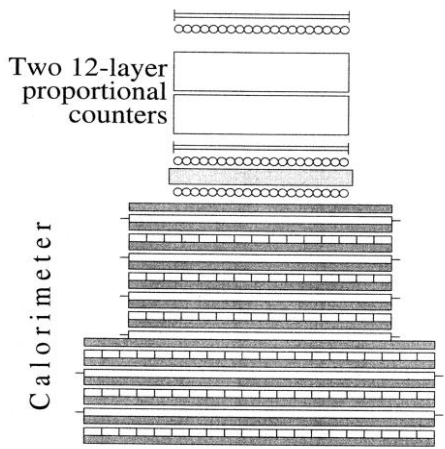

Fig. 2. The structure of Aragats experiment.

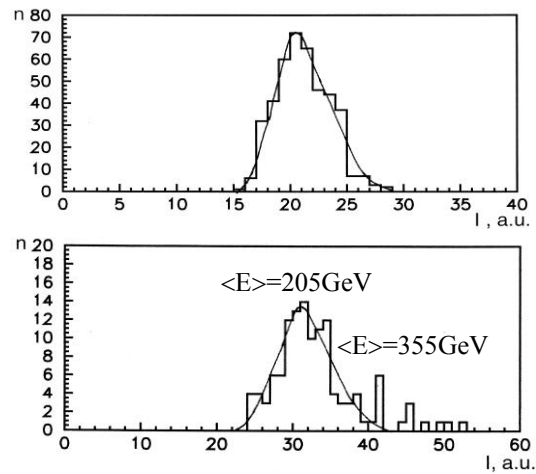

Fig. 3. Top: ionization spectrum of 0.6 GeV muon used for calibration. Bottom: ionization spectrum of hadrons. 14 events with ionization outside superimposed calibration curve have much higher average energy.

## 3. ATIC-2 single charge particle ionization analysis

The Advanced Thin Ionization Calorimeter (ATIC) [10] was designed to measure the composition and energy spectra of Z=1 to 28 cosmic rays over the energy range ~10 GeV to 100 TeV. The structure of ATIC device is shown on Fig. 4. First time it was launched as a long duration balloon test flight on December 28, 2000 from McMurdo, Antarctica. The second science flight of ATIC started on December 29, 2002 also from McMurdo and was completed on January 18, 2003 [11]. We tried to analyze the small region of the charge (ionization) distribution from zero to alpha particles, where FCP can be found. Unfortunately only one layer of silicon pad detectors was used in ATIC silicon matrix [10]. It seriously limits the possibility to separate 2e/3 particles from protons and makes practically impossible 4e/3 particle separation.

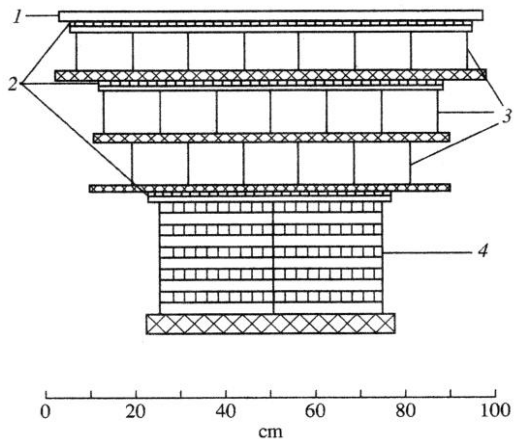

**Fig. 4. The layout of the ATIC device: (1) silicon matrix; (2) scintillator detectors; (3) carbon target; and (4) BGO ionization calorimeter.**

Fig. 5 shows a histogram of ionization distribution near the proton peak for all the particles with energy deposited in the BGO calorimeter above 50 GeV. The right smooth curve fitted to the proton peak is a normalized ATIC silicon matrix detector response to 1 m.i.p. particle. Left smooth curve is a distribution of the events, which do not belong to the proton distribution. One can see that relative half width of both curves approximately identical (~50%). The maximum on the right curve corresponds to the most probable ionization of the regular protons (0.87 arb. units). It agrees with the maximum of alpha particle ionization position (not shown), which is 4 times higher because of the 2 times higher charge. The maximum of the left curve with "additional" events corresponds to ionization of ~0.46 arb. units or ~0.53 of proton ionization. If we suppose that the "additional" events are real particles then their charge can be estimated as 0.73e, which is very near to 2e/3≅0.67e. The right plot on Fig. 5 presents the same ionization distribution but for central 60% of the silicon matrix area. One can see that the fraction of "additional" events relative to proton one is less there.

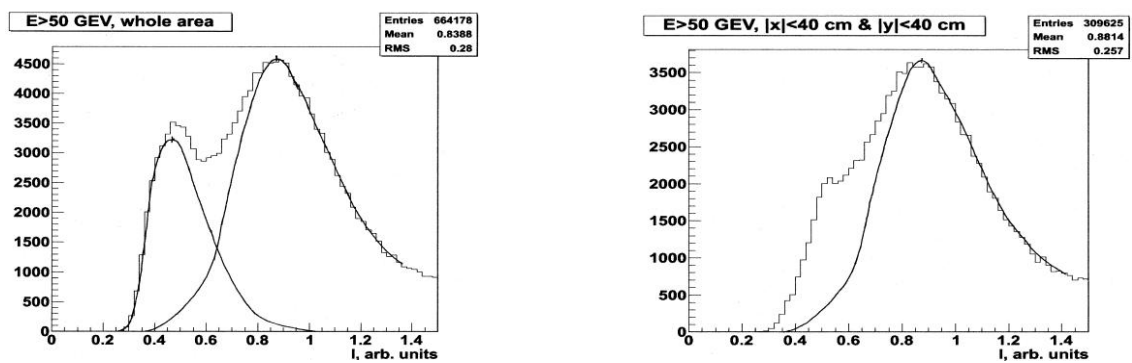

**Fig. 5. The ionization spectra for charged particles with energy E>50 GeV measured during the ATIC second flight. Left: whole area of the silicon matrix is used for analysis. Right: only central part of the silicon matrix (~60%) in use.**

Fig. 6 presents an angular distribution for the protons (above 50 GeV) (the left plot) and for the "additional" events (the right plot). One can see that the "additional" events angular distribution has a visible maximum around 35º. It can look very strange if device position is undetermined in space. But we should remember that the ATIC instrument was flying around the South Pole. Its axis has near to the permanent angle with the Earth axis about 15º. During the flight because of the Earth rotation the ATIC axis created relatively narrow cone, and only the limited area of the space has been observed. If certain anisotropy for some particle flux exists it can explain their different angular distribution.

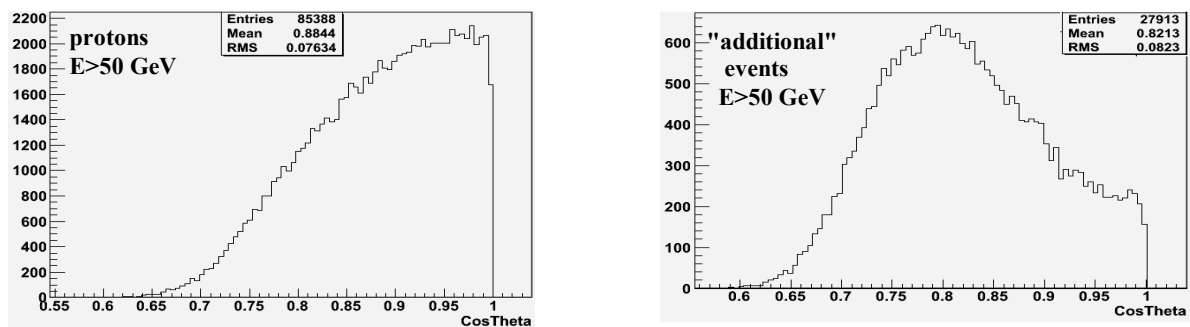

Fig. 6. Angular distributions for protons and "additional" events.

## 4. Conclusions

The hadrons with energy above 100 GeV is probably the most promising area to search for fractionally charged particles in Cosmic Rays.

To separate them reliably from regular particles we need 5-6 layers of silicon detectors with signal to noise ratio about 15, combined with a hadron calorimeter.

Rather simple apparatus like EM/Hadron Calorimeter [8] proposed for high energy electrons investigation can be simultaneously used for fractionally charged particles search and investigation.

## References

[1]  H. Faissner, M. Holder, K. Krisor, G. Mason, Z. Sawaf, and H. Umbach, Search for Fractionally Charged Particles in Showers of Low Density, Phys. Rev. Lett 24 (1970) 1357-1360.
[2]  A. F. Clark, R. D. Ernst, H. F. Finn, G. G. Griffin, N. E. Hansen, D. E. Smith, W. M. Powell, Cloud-Chamber Quark Quest-Negative Results, Phys. Rev. Lett 27 (1971) 51-55.
[3]  W. E. Hazen, Cloud-chamber search for 1/3 e and 2/3 e quarks in air showers, Phys. Rev. Lett 26 (1971) 582-583.
[4]  F. Ashton, A. J. Saleh, Search for e/3 Quarks in Regions of EAS of Local Electron Density > 500 m-2, in: the 14th International Cosmic Ray Conference, 1975, vol. 7, p. 2467.
[5]  F. Ashton, R. B. Coats, G. N. Kelly, D. A. Simpson, N. I. Smith, T. Takahashi, A search for relativistic quarks in the cosmic radiation, Journal of Physics A: Mathematical and General, vol. 1 (1968) 569-577.
[6]  E. P. Krider, T. Bowen, R. M. Kalbach, Search for charge 1/3e and 2/3e quarks in cosmic radiation near sea level, Physical Review D, vol. 1 (1970) 835-840.
[7]  W. T. Beauchamp, T. Bowen, A. J. Cox, R. M. Kalbach, Search for 1/3e and 2/3e charged quarks in the cosmic radiation at 2750-m altitude, Physical Review D, vol. 6 (1972) 1203-1211.
[8]  G. L. Bashindzhahyan, N. A. Korotkova, N. B. Sinev and L. G. Tkatchev, The EM/hadron calorimeter for high energy cosmic ray electrons investigation, in: the 31st International Cosmic Ray Conference, 2009.

[9] G. L. Bashindzhagyan, L. I. Sarycheva, N. B. Sinev, Search for new particles among the single cosmic ray hadrons, in: the 16th International Cosmic Ray Conference, 1979, vol. 6, p. 143.

[10] J. H. Adams Jr. for the ATIC collaboration, Preliminary results from the first flight of ATIC: The silicon matrix, in: the 27th International Cosmic Ray Conference, 2001, p. 2127.

[11] J. P. Wefel, J. H. Adams, H. S. Ahn, G. L. Bashindzhagyan, K. E. Batkov et al., The ATIC Science Flight in 2002-03: Description and Preliminary Results, in: the 27th International Cosmic Ray Conference, 2003, p. 1849.